\documentclass[12pt,preprint]{aastex}

\shorttitle{Enhanced X-ray and Infrared emission from AXP 1E 2259+58}
\shortauthors{Ertan $\&$ \c{C}al{\i}\c{s}kan  2005}

\begin{document}


\title{Optical and Infrared Emission from the AXPs and SGRs}


\author{
{\"{U}. Ertan\altaffilmark{1} $\&$ \c{S}. \c{C}al{\i}\c{s}kan \altaffilmark{2}}}


\altaffiltext{1}{Sabanc\i\ University, 34956, Orhanl\i\, Tuzla, \.Istanbul, Turkey}
\altaffiltext{2}{Bo\u{g}azi\c{c}i University, 34342, Bebek, \.Istanbul, Turkey} 

\begin{abstract} 

We show that the irradiated accretion disk model can account for all the optical and infrared observations of the anomalous X-ray pulsars in the persistent state. Model fits do not constrain the outer disk radii, while placing an upper limit to the inner disk radii, and thus to the strength of the dipole component of the stellar magnetic field. While magnetar fields ($B_* > 10^{14}$ G) in higher multipoles are compatible with the irradiated disk model, magnetic dipole components of magnetar strength are not consistent with optical data.       
 
\end{abstract}

\def\la{\raise.5ex\hbox{$<$}\kern-.8em\lower 1mm\hbox{$\sim$}}
\def\ga{\raise.5ex\hbox{$>$}\kern-.8em\lower 1mm\hbox{$\sim$}}
\def\be{\begin{equation}}
\def\ee{\end{equation}}  
\def\ba{\begin{eqnarray}}
\def\ea{\end{eqnarray}}  
\def\be{\begin{equation}}
\def\ee{\end{equation}}  
\def\ba{\begin{eqnarray}}
\def\ea{\end{eqnarray}}  
\def\m{\mathrm}
\def\d{\partial}
\def\R{\right}
\def\L{\left}
\def\a{\alpha}
\def\Mdot*{\dot{M}_*}
\def\Mdotin{\dot{M}_{\mathrm{in}}}
\def\Mdot{\dot{M}}
\def\Lin{L_{\mathrm{in}}}
\def\Rin{R_{\mathrm{in}}}
\def\Rout{R_{\mathrm{out}}}
\def\Rout{R_{\mathrm{out}}}
\def\Ldisk{L_{\mathrm{disk}}}
\def\Lx{L_{\mathrm{x}}}
\def\dEb{\delta E_{\mathrm{burst}}}
\def\dEx{\delta E_{\mathrm{x}}}
\def\Bb{\beta_{\mathrm{b}}}
\def\Be{\beta_{\mathrm{e}}}
\def\Rc{\R_{\mathrm{c}}}
\def\dMin{\delta M_{\mathrm{in}}}
\def\dM*{\delta M_*}
\def\Teff{T_{\mathrm{eff}}}
\def\Tirr{T_{\mathrm{irr}}}
\def\Firr{F_{\mathrm{irr}}}
\def\Av{A_{\mathrm{V}}}
\def\p{\propto}
\def\m{\mathrm}
\def\Ks{K_{\mathrm{s}}}




\section{Introduction}

Anomalous X-ray Pulsars (AXPs) and Soft Gamma-ray Repeaters (SGRs) constitute a special class of neutron star systems (Mereghetti et al. 2002; Hurley 2000; 
Woods $\&$ Thompson 2004). They are identified mainly through their X-ray luminosities 
($\Lx \sim 10^{34} - 10^{36}$ erg s$^{-1}$), which are orders of magnitude higher than their rotational powers $\dot{E}_{rot}= I \Omega \dot{\Omega}$. Their spin periods are clustered to a very narrow range ($5 - 12$ s). All five known SGRs and two of the eight known AXPs show repetitive, short (\la 1s) super-Eddington bursts with luminosities up to 
$10^{42}$ erg s$^{-1}$.  Three giant flares with peak luminosities 
$L_{\m{p}}> 10^{44}$ erg s$^{-1}$ 
and durations of a few minutes  were observed from three different SGRs 
(Mazets et al. 1979; Mazets et al. 1999; Hurley et al. 1999; Palmer et al. 2005).  

The short time scales and the  super-Eddington luminosities of the soft gamma-ray bursts strongly indicate a magnetar mechanism for these bursts. 
Magnetar models (Duncan $\&$ Thompson 1992; Thompson $\&$ Duncan 1995; 
Thompson $\&$ Duncan 1996) adopt strong  magnetic fields with magnitude $B_* > 10^{14}$ G on the stellar surface to explain the burst energetics.
Are the burst energies stored in the dipole component or the higher multipoles of the magnetic field of the neutron star?   
In the current magnetar models, the dipole component of the magnetic field must be of magnetar strength  to account for the spin down properties of the AXPs and SGRs. In these models, persistent X-ray luminosities are explained by the magnetic field decay, 
while the magnetic dipole torque is taken as the  mechanism responsible for the spin down rates of the sources.

In the alternative fallback disk model (Chatterjee, Hernquist, 
$\&$ Narayan 2000; Alpar 2001), the source of the X-rays is accretion onto the neutron star, while the optical/IR light originates from the accretion disk.   Rotational evolution of the neutron star is determined by the interaction between the disk and the magnetosphere of  the neutron star 
($B_*\sim 10^{12}- 10^{13}$ G).  Fallback disk models can account for the period clustering of AXPs and SGRs as the natural outcome of disk-magnetosphere interaction during their lifetimes (Alpar 2001; Ek\c{s}i $\&$ Alpar 2003). These models are consistent with magnetar fields on the neutron star provided that these fields are in higher  multipole components of the magnetic field. As higher multipole fields rapidly decrease with increasing radial distance (as $r^{-5}$ for quadrupole component), it is the dipole component of the magnetic field which determines the interaction and the angular momentum transfer between the disk and the neutron star.    
The strength of magnetic dipole field in order to explain the period clustering of the AXPs and SGRs over their $\Mdot$ history is 
$B_* \sim 10^{12} - 10^{13}$ G (Alpar 2001; Ek\c{s}i $\&$ Alpar 2003).

Fallback disk models can also explain the enhancements observed in the persistent luminosities of SGRs and AXPs.  
The X-ray enhancement of the SGR 1900+14 following its giant flare can be explained by the relaxation of a disk which has been pushed back by a preceding burst (Ertan $\&$ Alpar 2003). The same model with similar disk parameters can also reproduce the correlated X-ray and IR enhancement of AXP 2259+58, which lasted for $\sim$ 1.5 years, if this is triggered by a burst, with a burst energy estimated to have remained under the detection limits (Ertan, G\"{o}\u{g}\"{u}\c{s} $\&$ Alpar 2006). 

The suggestion of fallback disks has motivated observational searches for disk emission in the optical and IR bands, and resulted in various constraints on the models.   
Some of the AXPs were observed in more than one IR band
(Hulleman et al. 2001; Israel et al. 2002; Wang $\&$
Chakrabarty 2002; Kaspi et al. 2003; Israel et al. 2003; Hulleman et al. 2004;
Israel et al. 2004; Tam et al. 2004; Morii et al. 2005; Durant $\&$ van Kerkwijk 2006b). 
AXP 4U 0142+61 is the source with most extended observations, as it was 
also observed in the optical R and V bands 
(Hulleman et al. 2000; Hulleman et al. 2004; Dhillon et al. 2005), and recently in mid-IR bands with the 
{\it SPITZER} observatory (Wang et al. 2006). 
Discovery of modulation in the R band luminosity of 4U 0142+61 at the neutron star's rotation period P=8.7 s, with a pulsed fraction   
27 \% (Kern $\&$ Mertin 2002; Dhillon et al. 2005), is particularly significant.  This fraction is much higher than the pulsed fraction of 
the X-ray luminosity of this source, indicating that the origin of the pulsed optical emission is not likely to be the reprocessed X-rays by the disk. 
Magnetospheric models for these pulsations can be built with either a 
dipole magnetar field or within a disk-star dynamo model (Cheng $\&$ Ruderman
1991), in which a magnetospheric pulsar activity is sustained by a stellar dipole field of $\sim 10^{12}$ G and a disk protruding within the magnetosphere.  Ertan $\&$ Cheng (2004) showed that this pulsed optical component of the AXP 4U 0142+61 can be explained by both types of magnetospheric models. Thus the presence of strong optical pulsations from the magnetosphere does not rule out the possibility of a fallback disk together with $10^{12}-10^{13}$ G surface dipole magnetic field.

In the present work, we concentrate on the unpulsed optical/IR emission from the AXPs and SGRs in their persistent states, and test the expectations of the irradiated accretion disk model through the observations in different optical/IR energy bands (V, R, I, J, H, K, and K$_s$). The optical/IR emission expected from the irradiated fallback disks was first computed and discussed by Perna, Hernquist \& Narayan  (2000) and Hulleman, van Kerkwijk \& Kulkarni (2000). Using similar irradiation strengths, Perna et al. (2000) and Hulleman et al. (2000) found similar optical fluxes that remain well beyond those indicated by the observations of AXP 4U 0142+61 and AXP 1E 2259+586. To explain this result, Perna et al. (2000) suggested that the inner disk regions could be cut by an advection dominated flow, while Hulleman et al. (2000) concluded that the then existing optical data of the AXP 4U 0142+61 (in I,R,V bands) can only be accounted for by an extremely small outer disk radius, around a few $\times 10^9$ cm. In the present work, we show that the optical/IR data of the AXPs can be explained by the irradiated accretion disk model without any implausible constraints on the outer and inner disk radii. The main reason for the difference between our results and those of earlier works is that both Hulleman et al. (2000) and Perna et al. (2000)  assumed a particular irradiation strength, while we keep it as a free parameter, to address the broad band full data set. This approach is supported  by the observations of the low mass X-ray binaries (LMXBs) which indicate varying irradiation strengths. Furthermore, model fits are sensitive to the interstellar reddening parameter $\Av$, which was  estimated to be between 2.6 and 5.1 for 4U 0142+61 (Hulleman et al. 2004). For this source, we obtain the best model fit with $\Av=3.5$ which turned out to be consistent with the recent result $\Av=3.5\pm 0.4$ (Durant \& van Kerkwijk 2006a). On the other hand, the test disk model with unreasonably small outer disk radius requires $\Av=5.4$ (Hulleman et al. 2000). In this model, optical emission comes from the outer disk, while in our model it is the inner regions of an extended disk that emits substantially through the optical bands (see \S\ 3 for further discussion). We give the details of the disk model in \S\ 2. We discuss our results in \S\ 3, and summarize the conclusions in \S\ 4.

\section{Optical/IR Emission from the Irradiated Disk}

Model fits to the X-ray and IR enhancement data (Kaspi et al. 2003) of AXP 1E 2259+586 favor the irradiated disk model, though they do not exclude the nonirradiated thin disk model (Ertan, G\"{o}\u{g}\"{u}\c{s} $\&$ Alpar 2006). We start by assuming that the AXP disks are irradiated and include the irradiation strength as a free parameter through our calculations. 

When the disk is irradiated by the X-rays from the neutron star, both the intrinsic dissipation and the irradiation flux should be taken into account in calculations of the disk blackbody emission. A steady disk model is a good approximation for the present evolution of the AXP and SGR disks in their persistent states. For a steady thin disk, the intrinsic dissipation can be written as      
$D =(3/8 \pi)(G M \Mdot/R^3)$ (see e.g. Frank et al. 2002) 
where $\Mdot$ is the disk mass flow rate, $M$ is the mass of the neutron star and $R$ is the radial distance from the neutron star. In the absence of irradiation, the effective temperature $\Teff$ of the disk is proportional to 
$R^{-3/4}$ for a given $\Mdot$. For an irradiated disk, the irradiation flux  can be written as      
$\Firr=\sigma \Tirr^4 = (C \Mdot c^2)/(4 \pi R^2)$.   
(Shakura $\&$ Sunyaev 1973), where $c$ is the speed of light. Iraadiation parameter $C$ includes the effects of  the conversion efficiency of the 
accretion into X-rays, disk geometry and the albedo of the disk face.    
Irradiation temperature
$\Tirr= (\Firr/\sigma)^{1/4}$ is proportional to $R^{-1/2}$. 
For small radii, dissipation is the dominant source of the disk emission.  At a critical radius $R_c$, the irradiation flux becomes equal to the dissipation rate, and beyond $R_c$, the disk emission is supported  mainly by reprocessed X-rays. Equating $\Firr$ to $D$, the critical radius is found to be     
$R_c=(3GM_*)/(2 C c^2)\simeq  (10^{-4}/C) 3\times 10^9$ cm.
The effective temperature profile of the disk can be obtained using 
$\sigma \Teff^4 = D + \Firr $
where $\sigma$ is the Stefan-Boltzmann constant. 

We adopt the observed magnitudes in the optical/IR bands, distances and the $N_H$ values given by  Woods $\&$ Thompson (2004) and references therein and convert the magnitudes to energy flux values. We calculate $\Av$ values using $N_H = 1.79\times 10^{21} \Av$ (Predehl $\&$ Schmitt 1995). 
To find the model disk flux in a given observational band, we integrate the calculated blackbody emissions of all radial grids radiating in this band. For comparison with data, we calculate the model disk fluxes along the optical/IR bands V, R, H, I, J, K and K$_s$.  For all sources we set $\cos i=1$ where $i$ is the angle between the disk normal and the line of sight of the observer. We equate the disk mass flow rate $\Mdot$ to the accretion rate onto the neutron star, thus assuming the mass loss due to the propeller effect is negligible. We first adjust $\Mdot$ to obtain the observed X-ray flux. Next, using this value of $\Mdot$ and taking the strength of the magnetic dipole field $B_*= 10^{12}$ G on the surface of the neutron star we calculate the Alfv$\acute{e}$n radius $R_\m{A}$ which we take to be the inner radius of the disk. Then, we look for a good fit to the overall available optical/IR data by adjusting the irradiation strength $C$  within the uncertainties discussed in \S\ 3. 

\section{Results and Discussion} 

Our results are summarized in Table 1.  For each source, the first column  gives the unabsorbed flux data obtained from the observed magnitudes and the estimated  $\Av$ values (see Table 1) given in  Woods  $\&$ Thompson (2004), and the second column gives the model fluxes. For the AXP 4U 0142+61, the range of reddening quoted in earlier literature is $2.6 < \Av < 5.1$ (Hulleman et al. 2004). We obtain a good fit with $\Av=3.5$. Table 1 shows that the irradiated steady disk model is in agreement with all the AXPs observed in the optical and IR bands. The parameters of the model for each source are given in Table 2.  

At present, AXP 4U 0142+61, which has been observed in five different optical/IR bands from K to V in the same X-ray luminosity regime, seems to be the best source to study the properties of AXPs in the persistent state.   
Earlier work by Hulleman et al. (2000) excluded the disk model for the AXP 4U 0142+61. They obtained an irradiation temperature profile by using a particular irradiation strength. The estimated optical flux for an extended disk 
with this irradiation efficiency remains above the optical data points of the AXP 4U 0142+61 (see Fig. 3  in Hulleman et al. 2000).
Considering the possibility that the optical flux might originate from the outermost disk region,  Hulleman et al. (2000) tried  to fit the then observed three data points in the I, R and V bands to the Rayleigh-Jeans tail of a blackbody spectrum with the extinction parameter $\Av=5.4$. This placed an upper limit to the outer disk radius which is too small for a realistic disk. 
The key factor in the difference between the earlier results and our recent results is the irradiation efficiency, which we allow to vary in conjunction with $\Av$, to provide the best fit to the current broad band data.   
We note that the irradiation efficiency indicated by the observations of the low mass X-ray binaries varies from source to source. Even for the same source, the ratio of the irradiation flux to the X-ray flux may change with accretion rate (de Jong et al. 1996;  Dubus et al. 1999; Ertan $\&$ Alpar 2002). Taking these into account, we keep the irradiation efficiency as a free parameter for our model fits. With the parameters given in Table 2, the irradiated disk model can account for the optical/IR data of this source  without setting any stringent constraints on the inner or outer disk radii. In our model, the optical luminosity is radiated from the inner disk, while longer wavelength IR emission comes from larger radii.  A more detailed analysis of the AXP 4U 0142+61 with the new detections in the mid-IR $SPITZER$ bands confirms the results here (Ertan et al. 2006).       
The irradiation parameter $C$ obtained from our model fits turned out to be in the range ($10^{-4} < C < 10^{-3}$) estimated from the observations of LMXBs and the disk stability analyses of the soft X-ray transients (de Jong et al. 1996;  Dubus et al. 1999; Ertan $\&$ Alpar 2002). Within the critical radius $R_\m{c}$ given by Eq. 4, dissipation is the dominant heating mechanism. For the disk model of the AXP 4U 0142+61, $R_\m{c}\simeq 3\times 10^9$ cm and $\Rin = 1\times 10^9$ cm. The innermost disk emitting mostly in the UV bands also contributes to the optical emission. The radial distance at which the disk blackbody temperatures peak at the optical bands (R,V) is about $10^{10}$ cm.
at this radial distance,  
about 35 \% of of the optical radiation is due to dissipation, and the rets is due to irradiation. Peak temperatures of the IR bands from I to K$_\m{s}$ lie between $R\sim 2\times 10^{10}$ cm and 
$R\sim 1.5\times 10^{11}$ cm.    
    
There are several uncertainties related to the inner disk emission characteristics of the AXPs, which are not possible to address by the irradiated thin disk model. Fistly, emission properties of the innermost disk boundary interacting with the magnetosphere are not very clear. Secondly, the contributions from the  magnetospheric pulsed emission which is known to have a fraction about 27 \% in the R band for 4U 0142+61, is likely to be radiated from the other IR and optical bands as well. Relative amplitudes of these pulsed contributions radiated from different optical/IR bands are not known at present. Finally, there could be some X-ray shielding effects depending on the details of the geometry of the innermost disk regions, which could also affect the optical/IR emission properties of these sources. For all the AXPs that were detected in the optical/IR bands, optical and IR flux values of our models remain within about 30 \% of all the data points, which is a reasonable fit considering the uncertainties discussed above.  

For AXP J1708-40, Durant and van Kerkwijk (2006b) recently found that the previously reported IR data in $\Ks$, H, and J bands are likely to be a background star. They found another object within the positional error cycle and argued that this second object is more likely to be the IR counterpart to the AXP J1708-40. For this source, we adopt the IR ($\Ks$, H, J) data set reported by  Durant and van Kerkwijk (2006b).

For the persistent state of the AXP 1E 2259+586, we use the preenhancement data  (Hulleman et al. 2001). This source was detected in  $\Ks$ band and there are upper limits for I and R bands. Our model flux values are three and ten times below the upper limits reported for I and R bands respectively. 

AXP 1E 1048-59 was detected in  $\Ks$, H and I bands (Wang $\&$ Chakrabarty 2002). Observed X-ray flux from this source between December 2000 to January 2003 show a variation within a factor of 5 (Mereghetti et al. 2004). We use the X-ray flux obtained from  the nearest X-ray observation to the date of the IR observations. 

AXP 1E 1841 was detected only in the $\Ks$ band, and there is a high upper limit in the R band (Watcher et al. 2004). Model estimates in other optical/IR bands for this source (and the other AXPs) can be tested by future optical and IR  observations.      

Since there are no detections in short wavelength optical bands for the AXPs (except for AXP 4U 0142+61), model fits are not sensitive to the chosen inner disk radii. We equate the inner disk radii to the Alfv$\acute{\m{e}}$n
radii (Table 2) corresponding to a magnetic field with magnitude $B_* = 10^{12}$ G on the stellar surface and the accretion rates derived from the estimated X-ray luminosities (see Table 1 for references).  
For the AXP 4U 0142+61, optical data in R and V bands provide a constraint for the inner disk radius, and thereby for the strength of the magnetic dipole field of this source (see \S\ 4).

\section{Conclusion}

We have shown that the optical, infrared and X-ray observations of the AXPs in their persistent states can be explained with irradiated disk models. Among the AXPs, 4U 0142+61 is currently the only source which provides  an upper limit for the inner disk radius through its optical (R,V) data. For the best model fit for this source, which we have obtained with  $\Av =3.5$,  the model inner disk radius ($\sim 10^9$ cm) is around the Alfv$\acute{e}$n radius for the accretion rate, estimated from the X-ray luminosity, together with a dipole magnetic field strength  $B_* \simeq 10^{12}$ G on the neutron star surface. Nevertheless, it is possible to obtain reasonable fits by increasing the inner disk radius and decreasing the reddening accordingly. For $\Av=2.6$, the minimum value of the reddening in the range 
$2.6 < \Av < 5.1$ (Hulleman et al. 2004), we obtain the best fit with 
$\Rin \simeq 8\times 10^9$ cm which corresponds to the maximum reasonable dipole field strength  
$B_*\simeq 4\times 10^{13}$ G on the pole and half of this on the equator of the neutron star. We note that these limits could be increased depending on the amount of possible mass loss due to propeller effect and/or on how much the inner disk radius penetrates inside the Alfv$\acute{e}$n radius (see Ertan et al. (2006) for a detailed discussion for 4U 0142+61). On the other  hand, even including these possibilities, very recent analysis concluding $\Av=3.5\pm 0.4$ (Durant \& van Kerkwijk 2006a) implies surface dipole magnetic field strengths less than about $10^{13}$ G.
While the magnetar fields ($B_* > 10^{14}$ G) in multipoles are compatible with this picture, optical (R, V) data excludes a hybrid model involving a disk surrounding a  magnetar dipole field. In the latter case, inner disk regions emitting in the optical would be truncated by the magnetar dipole field. High magnetic fields in multipoles, on the other hand, decrease rapidly with increasing radial distance, and do not affect the disk magnetosphere interaction. 

Uncertainties in the source distances, require modification of the model  $\Mdot$ values, and thereby the disk temperature profiles. For the AXP 4U 0142+61, new fits with similar quality can be obtained by slightly modifying the model inner disk radius and the irradiation parameter $C$ as long as such distance corrections remain within a factor of $\sim 3$. In this case, surface dipole magnetic field should also be recalculated consistently with the newly obtained  $\Mdot$ and Alfv$\acute{e}$n radius, $r_{A}\propto \Mdot^{-2/7} B_*^{4/7}$. 
For the other AXPs, which were observed only in the IR bands, model fits at present are not sensitive to the inner disk radius, and in the case of distance corrections, similar fits can be obtained only by changing the irradiation strength. This is because IR radiation is emitted from the irradiation dominated outer disk regions, and therefore $\Mdot$ corrections do not change the relative amplitudes of the fluxes in different IR bands. Future detections of these sources in the optical bands can constrain the inner disk radii together with the surface dipole field strengths.

On the other hand, existing IR data of the AXPs, including recent observations of 4U 0142+61 by {\it SPITZER} in 4.5 $\mu$m and 8 $\mu$m bands (Wang et al. 2006), do not put an upper limit for the extension of the outer disk radius $\Rout$. The lower limit for $\Rout$ provided by the longest wavelength IR data of the AXP 4U 0142+61 is around $10^{12}$ cm.   
Further observations in the longer wavelength infrared bands by $SPITZER$ space telescope will provide valuable information about the structure and possibly the extension of the fallback disks around these systems. As a final remark, some  AXPs and SGRs which are under the detection limits in some of the optical and IR bands could be observed in these bands if they exhibit phases of enhanced emission, as observed in the SGR 1900+14 and the AXP 1E2259+586.

\acknowledgments

We thank Ali Alpar, Ersin G\"{o}\u{g}\"{u}\c{s} and Yavuz Ek\c{s}i for valuable  comments on the manuscript.  
We acknowledge support from the Astrophysics and Space
Forum at Sabanc{\i} University. 
S.\c{C}. acknowledges support from the FP6 Marie Curie Reintegration Grant, INDAM.


\clearpage

\begin{deluxetable}{l|cc|cc|cc|cc|cc}
\rotate
\tablecaption{}
\tablewidth{0pt}
\startdata
\tableline\tableline
&\multicolumn{2}{c|}{J1708-40}&\multicolumn{2}{c|}{1E 2259+58} 
&\multicolumn{2}{c|}{4U 0142+61}
&\multicolumn{2}{c|}{1E 1841-045}
&\multicolumn{2}{c}{1E 1048-59}
\\
\tableline
&\multicolumn{2}{c|}
{Flux}&\multicolumn{2}{c|}
{Flux }&\multicolumn{2}{c|}
{Flux }&\multicolumn{2}{c|}
{Flux }&\multicolumn{2}{c}
{Flux }
\\
&\multicolumn{2}{c|}
{($10^{-15}$erg s$^{-1}$ cm$^{-2}$) }&\multicolumn{2}{c|}
{($10^{-15}$erg s$^{-1}$ cm$^{-2}$)}&\multicolumn{2}{c|}
{($10^{-15}$erg s$^{-1}$ cm$^{-2}$)}&\multicolumn{2}{c|}
{($10^{-15}$erg s$^{-1}$ cm$^{-2}$) }&\multicolumn{2}{c}
{($10^{-15}$erg s$^{-1}$ cm$^{-2}$) }
\\
\tableline
Band&Data&Model &Data&Model &Data&Model&Data&Model&Data&Model  
\\
&$(\Av=7.8)$&&$(\Av=6.1)$&&$(\Av=3.5)$&&$(\Av=8.4)$&&$(\Av=5.6)$&
\\
\tableline
K$_s$&49&44&3.7 &3.6 &14 &14  &68 &68 &29&22  
\\
\tableline
K  &  & 54&     &4.5 &18 &18  &  & 84 & &27  
\\
\tableline
H&51 &57 &     &4.8 & 19  &19  &  &89 &22&28  
\\
\tableline
J&50 &53 &     &4.4 & 14 &18  &  &83 &33&26  
\\
\tableline
I&   &56&$<$15  &4.4 &18 &21  &  &88 & & 24 
\\
\tableline
R&   &65 &$<$42 &4.5 &19 &25  &$<3.8\times10^5$&100 & &26  
\\
\tableline
V&   &48 &      &3.0 &28 &20  &  &76 & &18  
\\
\tableline

\enddata
\tablecomments{The data flux values were calculated by using the magnitudes and $\Av$ values given in the references below. For the AXP 4U 0142+61 plausible range for reddening is 2.6$ < \Av < 5.1$ (Hulleman et al. 2004); the data of this source here correspond to $\Av$=3.5.  REFERENCES: (J1708-40) Durant \& van Kerkwijk 2006b, Rea et al. 2003; (1E 2259+586) Hulleman et al. 2001, Woods et al.  2004; (4U 0142+61) Hulleman et al. 2000, 2004, Patel et al. 2003, Morii et al. 2005; (1E 1841-45) Wachter et al. 2004, Morii et al. 2003; (1E 1048-59) Wang $\&$ Chakrabarty 2002, Mereghetti et al. 2004 }
\end{deluxetable}

\clearpage

\begin{table}

\begin{center}
\caption{The parameters of the irradiated disk model which gives the optical/IR flux values seen in Table 1. For all the sources, we set $\cos i=1$ where $i$ is the inclination angle between the disk normal and the line of sight of the observer, and we take the outer disk radius $\Rout=5\times 10^{12}$ cm. See Section 3 for details.  }

\begin{tabular}{l|c||c||c||c|c}
\tableline\tableline
&1RXS J1708-40&1E 2259+58 &4U 0142+61&1E 1841-045
&1E 1048-59
\\
\hline
$\Rin$ (cm)&$1.2\times 10^9 $& $2.3\times 10^9$
&$1.0\times 10^9$&$1.3\times 10^9$&
$3.3\times 10^9$   
\\
\hline
$C$&$5.0\times 10^{-4}$&$ 1.6\times 10^{-4}$
&$1.0\times 10^{-4}$&$7.2\times 10^{-4}$&
$7.0\times 10^{-4}$    
\\
\hline
$d$(kpc)& 5& 3& 3&7 &3   
\\
\hline
$\Mdot$ (g s$^{-1})$&$1.0\times 10^{15}$&$9.1\times 10^{13}$&$4.8\times 10^{14}$&$2.2\times 10^{15}$&$ 1.3\times 10^{14}$
\\
\tableline

\end{tabular}
\end{center}
\end{table}

\end{document}